\documentclass[aps,epsfig,twocolumn]{revtex4}
\usepackage{amsmath}
\usepackage{epsfig}
\usepackage{graphics}
\begin{document}
\title{Shear viscosity and damping for a Fermi gas in the unitarity limit}

\author{G.\ M.\ Bruun}


\author{H.\ Smith}

\affiliation{ Niels Bohr Institute, Universitetsparken 5, DK-2100
Copenhagen \O, Denmark.}

\date{\today{}}

\begin{abstract}
The shear viscosity of a two-component Fermi gas in the normal
phase is calculated as a  function of temperature in the unitarity
limit, taking into account strong-coupling effects that give rise
to a pseudogap in the  spectral density for  single-particle
excitations. The results indicate that recent measurements of the
damping of collective modes in trapped atomic clouds  can be
understood in terms of hydrodynamics, with a decay rate given by
the viscosity integrated over an effective volume of the cloud.
\end{abstract}

\maketitle

Pacs Numbers: 03.75.Ss, 05.30.Fk,51.20.+d


\section{Introduction}

Strongly interacting Fermi systems play a central role in physics
over a vast range of energies, from  cold atoms over condensed
matter systems to quark-gluon plasmas. For atomic gases, the
regime of strong interaction  is reached by the use of Feshbach
resonances at which  the scattering length diverges. This
so-called unitarity limit has been studied experimentally through
the expansion of a two-component Fermi gas~\cite{Ohara} and by
measuring its collective modes~\cite{Grimm,Kinast}. These
experiments indicate that under certain conditions the dynamic
properties of atomic gases in the unitarity limit are well
described by hydrodynamics, both in the superfluid and in the
normal phase. Related results have been reported for the strongly
interacting quark-gluon plasma produced in heavy-ion collisions at
RHIC~\cite{Xu,Teaney}.

In this paper we shall carry out a quantitative analysis of the
hydrodynamic damping for the normal phase in the unitarity limit
and compare our results to the measured rate of decay of the
collective modes~\cite{Grimm,Kinast}. In the hydrodynamic limit,
the rate of decay is related to the shear viscosity integrated
over the volume of the trapped atomic cloud. As we shall see,
under the given experimental conditions, hydrodynamics applies
only in a limited temperature region above the superfluid
transition temperature $T_c$. Furthermore, hydrodynamics
necessarily fails in the outer parts of the atomic cloud, where
the density is low and the mean free path therefore long. Since
for a classical gas the viscosity is independent of density, one
must introduce an explicit cut-off in the spatial integration of
the viscosity, as shown in Ref.\ \cite{Kavoulakis}.

In the unitarity limit, we can use a dimensional
argument~\cite{Gelman} to write $\eta$ as
\begin{equation}
\eta=n\hbar\alpha(T/T_F). \label{alphadef}
\end{equation}
Here $\alpha$ is a dimensionless quantity which can only depend on
temperature through $T/T_F$. The Fermi temperature is $T_F=k_F^2/2m$, where
$k_F$ is the magnitude of the Fermi momentum given by $k_F=(3\pi^2n)^{2/3}$
with $n$ being the density
of the gas (with the exception of (\ref{alphadef}), we use units such that $\hbar=k_B=1$).
 Our aim is to obtain an approximate expression for the
universal function $\alpha(T/T_F)$ which will allow us to compare
theory with experiment.

The present work is a continuation of two previous papers
\cite{Massignan, BruunSmith}, in the following  referred to as I
and II respectively, on the damping of collective modes in Fermi
gases. Before we describe the results of our present calculation
we shall therefore summarize the approach taken in I and II, and
indicate their limits of applicability.

In I we employed a Boltzmann equation for the fermion distribution
function $f({\bf r},{\bf p}, t)$, taking into account the
dependence of the scattering cross section on the energy in the
relative motion of two particles, as well as the effect of the
mean field in the streaming terms of the Boltzmann equation. The
collective mode frequencies  were calculated by taking moments of
the Boltzmann equation, which introduced the spatially-averaged
viscous relaxation rate as the effective collision rate entering
the imaginary part of the (complex) mode frequencies. At low
temperatures, well below the Fermi temperature, the validity of
this approach is restricted to the limit of weak coupling,
$k_F|a|\ll 1$, where $a$ is the scattering length and $k_F$ is the
magnitude of the Fermi wave vector. In this limit the method is
accurate  within a few percent, since the viscous relaxation rate
used in I is obtained from a trial function which yields
viscosities that differ by only a few percent from those obtained
from  exact solutions to the Boltzmann equation at both low and
high temperatures.  At temperatures well above the Fermi
temperature the Boltzmann approach used in I is accurate for any
value of $a$, including the unitarity limit in which the cross
section is proportional to the inverse of the energy in the
relative motion. We shall demonstrate this in
detail in App.\ \ref{hightemp}.

When $k_F|a|$ becomes comparable to or larger than unity, one
enters the regime of strong coupling, in which perturbative
approximation schemes such as the Boltzmann approach can no longer
be trusted at temperatures comparable to or less than the Fermi
temperature. Progress in understanding the viscosity of such
strongly coupled Fermi systems must necessarily rely on an
interplay between experiment and theory, since there is no small
parameter available for a perturbation expansion that could yield
firm theoretical predictions. For an atomic gas close to a
Feshbach resonance  we explored in II the influence of the medium
on the scattering cross section, which in I was taken to be its
value in vacuum. Due to Fermi blocking of the pair states into
which the molecular state can decay, the lifetime of the resonant
state was found to be significantly increased, leading to a
corresponding increase in scattering rate (and hence a decrease in
viscosity) close to the superfluid transition temperature $T_c$.
For a uniform gas the calculated viscosity just above $T_c$ was
found to be reduced by the factor 7.5 compared to the value
obtained in I by  use of the vacuum scattering matrix. For a
trapped gas the difference was less pronounced: the thermal
relaxation rate, which is closely related to the inverse
viscosity, was found near $T_c$ to be 3.6 times the value obtained
using the vacuum scattering matrix.

The assumption underlying the approach taken in II was that the
main effects of the interaction in the strong-coupling limit arose
through a modification of the scattering cross section, while
strong-coupling effects that lead to spectral broadening of
single-particle excitations were not taken into account. Put in
different terms, only the collision term in the kinetic equation was modified by taking into
account the medium effects mentioned above, while the streaming
terms were assumed to be unaffected by
interactions. In the present paper we abandon this assumption and
consider specifically the role of the pseudogap occurring in the
spectral function of single-particle excitations~\cite{Perali}.
The presence of the pseudogap in the normal phase influences the
Bragg spectrum observed when an atom absorbs a photon from one
laser beam and emits a photon into another, resulting in a change
of the energy and  momentum of the atom~\cite{BruunBaym}. In the
normal phase, at the unitarity limit, the pseudogap was found to
cause a significant suppression of the low-frequency Bragg
spectrum.

The use of a Boltzmann equation  as in I and II implicitly assumes
the existence of quasiparticles with a definite energy-momentum
relationship. When the spectral functions broaden, the
quasiparticles are less well-defined, and it therefore becomes
relevant to investigate the effect of this broadening on the
transport properties of the gas. Ideally, one should derive a
kinetic equation that takes all strong-coupling effects
systematically into account, but due to the lack of a small
parameter in the strong-coupling limit this would be far too
ambitious an undertaking. Our aim here is more modest: to compare
results for the viscosity in the presence and absence of spectral
broadening in order to gain insight into its quantitative
significance, and to use this together with the results of I and
II to construct an approximate formula for $\alpha$ that allows
for an explicit comparison with experiment.

A main result of the paper is the calculation in Sec.\ \ref{kubo} of the shear
viscosity from a Kubo formula, which allows one to take into
account the presence of the pseudogap in the spectral density of
states for single-particle excitations. Since our results are
based  on a ladder approximation to the self-energies, they cannot
be expected to be quantitatively accurate, but as we shall see,
our present results are quite close to those of paper II at low
temperatures. At high temperatures, however, the  Kubo expression
gives results that lie significantly below the classical result
obtained from the Boltzmann equation. Since the latter is
essentially exact for all values of $a$ in the classical limit, 
as we shall demonstrate
in App.\ \ref{hightemp}, we construct an approximate formula
for $\alpha$ which interpolates between the low-temperature result
and the exact high-temperature limit. This interpolation formula
is then used for comparison with experiment in Sec.\ \ref{exp},
where the decay rate is related to the viscosity integrated over
an effective volume of the trapped gas. The resulting agreement
with experiments \cite{Grimm,Kinast} that were carried out at two
very different frequencies indicates that the interpolation
formula provides a qualitatively correct picture of the physics
involved in the viscosity of a strongly interacting
Fermi gas. The calculations also illustrate how information on the
viscosity of strongly interacting Fermi gases can be extracted
from measurements of the damping of collective modes.

 \section{The Shear Viscosity} \label{kubo}

Consider the shear viscosity $\eta$ of a uniform,
two-component Fermi gas of atoms with mass $m$ in the normal phase. There is no
interaction between atoms in the same internal state whereas the
interaction between atoms in the two different internal states is
characterized by the s-wave scattering length $a$. Unitarity means
that we take $k_F|a|\rightarrow\infty$. The shear viscosity relates the momentum
current density $\Pi_{xy}$ to the gradient in  flow velocity
$u_x(y)$ according to $\Pi_{xy}=-\eta\partial u_x/\partial y$.

The Landau-Boltzmann approach assumes well-defined quasiparticle
excitations. However, with strong interactions
present the spectral functions may become significantly broadened
and the quasiparticles therefore ill-defined. Close to $T_c$ in the
normal phase, the spectral weight is found to be suppressed near the
Fermi surface resulting in a double-peak structure of the spectral
function~\cite{Perali,BruunBaym}. This suppression is referred to as
the presence of a pseudogap. In order to investigate the importance
of the pseudogap we turn to the Kubo formalism, which allows for a
treatment of these effects.

\begin{figure}
\includegraphics[width=0.9\columnwidth,
height=0.3\columnwidth,angle=0,clip=]{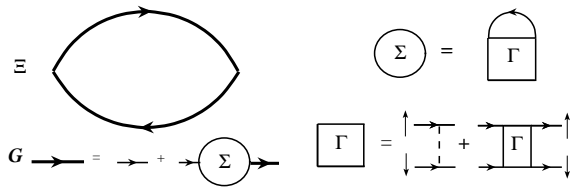} \caption{$\Xi$
and the atomic propagator $G$ in the ladder approximation for a broad
resonance. $\Sigma$ is the self energy and $\Gamma$ is the scattering matrix.} \label{FeynFig}
\end{figure}

The velocity field $u_x(y)$ gives rise to a perturbation
\begin{equation}
\hat{H}'=-m\int d^3r\,u_x(y)\hat{j}_x({\mathbf{r}})=
\frac{1}{i\omega}\frac{\partial u_x}{\partial y}\int
d^3r\,\hat{\Pi}_{xy} \label{Hpert}
\end{equation}
with $\hat{\mathbf{j}}$ being the current density operator and $\hat{\Pi}_{xy}$ the
momentum-current density operator. To obtain the
second equality in (\ref{Hpert}), we have used momentum conservation
$m\partial\hat{j}_i/\partial t=-\partial\hat{\Pi}_{ij}/\partial
r_j$ ($i$ and $j=x,y,z$) taking
$\hat{\mathbf{j}}({\mathbf{r}},t)=\exp(-i\omega
t)\hat{\mathbf{j}}({\mathbf{r}})$.
We can now calculate  $\Pi_{xy}$ induced by $\hat{H}'$ within linear
response. The shear viscosity is then obtained by taking the limit
$\omega\rightarrow 0$:
\begin{equation}
\eta=-\lim_{\omega\rightarrow 0}\rm{Im}\,\Xi(\omega)/\omega
\end{equation}
with
\begin{equation}
\Xi(\omega)=-i \int d^3rdt\,e^{i\omega t}
\theta(t)\langle[\hat{\Pi}_{xy} ({\mathbf{r}},t),\hat{\Pi}_{xy} (0,0)]\rangle.
\end{equation}
The approximation for $\Xi$ shown in Fig.\ \ref{FeynFig} yields
\begin{equation}
\eta=-\frac{1}{15m^2}\int \frac{d^3p}{(2\pi)^3}p^4\int
\frac{d\epsilon}{2\pi}A^2(p,\epsilon)
\frac{\partial f(\epsilon)}{\partial \epsilon}.
\label{ViscFormula}
\end{equation}
Here, $A(p,\epsilon)=-2{\rm Im}G_R(p,\epsilon)$ is the spectral
function for the atoms with $G_R$ being the retarded Green
function and $f(\epsilon)=[\exp(\epsilon/T)+1]^{-1}$. 
A relativistic version of (\ref{ViscFormula})  has been
used to calculate $\eta$ for quark-gluon plasmas using a
phenomenological ansatz for $A(p,\epsilon)$~\cite{Peshier}. Here,
we use a microscopic theory for $A(p,\epsilon)$ as explained
below.

To obtain $A(p,\epsilon)$, we calculate the Green function using a
multichannel BEC--BCS crossover theory based on the ladder
approximation for the thermodynamic potential and the atom self
energies. This is the minimal approach which includes the correct
two-body physics leading to the Feshbach
resonance~\cite{holland,griffin,BECBCSBruun}. The structure of the
theory is shown in Fig.\ \ref{FeynFig} and is described in detail
in Refs.\ \cite{BruunBaym,BECBCSBruun}. We take a broad resonance
for which the multichannel theory becomes equivalent to the
original single channel BEC--BCS crossover theory~\cite{Randeria}
for most observables.
 The spectral functions are found numerically to obey the sum rule
$\int A(p,\omega)d\omega=2\pi$ to a very good approximation. It
should be noted, however, that the approximation leading to
(\ref{ViscFormula}) is not conserving. To obtain a conserving
approximation, we need to solve an integral equation  for
$\Xi$~\cite{Baym}. However, the present analysis is already heavy
numerically since one needs to integrate over the atom self
energies which themselves are evaluated numerically; a conserving
approximation is thus beyond the scope of the present paper.

It is instructive to compare the Kubo approach with the
relaxation-time approximation to the Boltzmann equation, since
these yield identical results at high temperatures. In the
relaxation-time approximation, the collision integral of the
Boltzmann equation given in Eq.\ (13) of Ref.~\cite{Massignan}
becomes $I[f]=\delta f/\tau$, where $\delta f$ is the deviation of
the distribution function from local equilibrium. The relaxation
time $\tau(p)$ is obtained by setting the distribution functions
for particles in states other than $\bf p$ equal to their
equilibrium values,
\begin{gather}
\frac{1}{\tau(p)}=\int\frac{d^3p_1}{(2\pi)^3}\int
d\Omega\frac{d\sigma}{d\Omega}\frac{|{\mathbf{p}}-{\mathbf{p}}_1|}{m}\times\nonumber\\
\frac{f(\xi_{p_1})[1-f(\xi_{p'})][1-f(\xi_{p_1'})]}{1-f(\xi_p)}.
\label{relaxtime}
\end{gather}
Here $d\sigma/d\Omega$ is the differential cross section for the
scattering of two particles with incoming momenta ${\mathbf{p}}$
and ${\mathbf{p}}_1$ to outgoing momenta ${\mathbf{p}}'$ and
${\mathbf{p}}'_1$. The corresponding energies are $\xi_p=p^2/2m-\mu$ etc., 
and $\Omega$ is the solid angle
of $({\mathbf{p}}'-{\mathbf{p}}_1')/2$ with respect to
$({\mathbf{p}}-{\mathbf{p}}_1)/2$. It is straightforward to find
$\delta f$ from the Boltzmann equation and thus the momentum
current density from $\Pi_{xy}=2(2\pi)^{-3}\int
d^3p(p_xp_y/m)\delta f$, where the factor of $2$ is from the two
internal states. We obtain for the shear viscosity
\begin{equation}
\eta=-\frac{2}{15m^2}\int\frac{d^3p}{(2\pi)^3}p^4\frac{\partial
f}{\partial\xi_p}\tau(p). \label{etarelax}
\end{equation}
The factor $1/15$ arises from the angular average of $p_x^2p_y^2$.

The Kubo formula (\ref{ViscFormula}) reduces to the
relaxation-time approximation (\ref{etarelax}) in the limit of
weak interaction with narrow spectral function peaks. To see this,
put $A(p,\epsilon)=2\Gamma_p/[(\epsilon-\xi_p)^2+\Gamma_p^2]$ with
$\Gamma_p=-{\rm Im}\Sigma(p,\xi_p)$ the imaginary part of the atom
self energy. Using $A^2\rightarrow
2\pi\delta(\epsilon-\xi_p)/\Gamma_p$ for $\Gamma_p\rightarrow 0$,
(\ref{ViscFormula})  reduces to (\ref{etarelax}) with
$\tau^{-1}(p)=-2{\rm Im}\Sigma(p,\xi_p)$. It can furthermore be
shown that the ladder approximation for ${\rm Im}\Sigma(p,\xi_p)$
yields (\ref{relaxtime}) when medium effects are
ignored~\cite{BECBCSBruun}. The Kubo formula for the viscosity
(\ref{ViscFormula}) thus reduces to  (\ref{etarelax}) in the limit
of weak interactions.

\begin{figure}
\includegraphics[width=0.8\columnwidth,
height=0.6\columnwidth,angle=0,clip=]{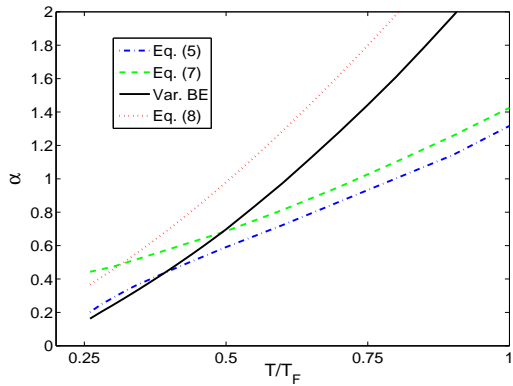} \caption{(Color
online) The viscosity in units of $n\hbar$ for $T\ge T_c$. The 
dashed-dotted
line is the Kubo formula (\ref{ViscFormula}),
 the dashed line is the Boltzmann equation result in the relaxation-time approximation (\ref{etarelax}),
the solid line is the variational solution of the Boltzmann
equation with the medium cross section~\cite{BruunSmith}, and the
dotted line is the high-$T$ result (\ref{etaclasslimit}).}
\label{etaFig}
\end{figure}

In Fig.\ \ref{etaFig}, we plot $\alpha(T/T_F)$ in (\ref{alphadef})
for a gas in the unitarity limit with $T\ge T_c$. For the
numerical calculations, we have chosen parameters corresponding to
a resonant interaction with $k_{\rm F}|a|=25\gg1$ and a negligible
effective range. The critical temperature is found from the
divergence of the scattering matrix at zero total momentum (the
Thouless criterion) yielding $T_c\approx 0.26T_{\rm F}$ in good
agreement with other BEC--BCS crossover
results~\cite{holland,Pieri}.
The Kubo result (\ref{ViscFormula}) approaches (\ref{etarelax})
for $T\gg T_F$. This is to be expected since medium effects are
negligible in the classical limit. For $T\gg T_F$, $\eta\propto
T^{3/2}$ which can be seen in the relaxation-time approximation
from (\ref{relaxtime})-(\ref{etarelax}) which give $\eta=nT\tau$
with $\tau\propto T^{1/2}$, resulting in $\alpha
=1.1(T/T_F)^{3/2}$. At low $T$, the difference between
(\ref{etarelax}) and (\ref{ViscFormula}) is significant; for
$T=T_c$, (\ref{etarelax}) yields $\alpha = 0.4$ whereas
(\ref{ViscFormula}) predicts $\alpha = 0.2$. This reduction is due
to strong-coupling medium effects. The imaginary part ${\rm
Im}\Sigma$  is large leading to a significant damping of the
quasiparticles. Close to $T_c$ a pseudogap opens up and the
effective density of states at the Fermi surface is suppressed,
leading to a reduction of the viscosity.

We also plot $\eta$ obtained from a variational solution to the
Boltzmann equation~\cite{Massignan, BruunSmith}, which yields a
lower bound on the viscosity, given by
\begin{equation}
\alpha=\frac{45\pi^{3/2}}{64\sqrt{2}}\left(\frac{T}{T_F}\right)^{3/2}=2.77\left(\frac{T}{T_F}\right)^{3/2}.
\label{etaclasslimit}
\end{equation}
This lower bound deviates by less than two percent from the exact
result, as we shall demonstrate in the Appendix, where we
investigate the leading correction to the lowest-order variational
result for a general form of the cross section. In the unitarity
limit the leading correction is found to increase the lower bound
(\ref{etaclasslimit}) by 3/190 or 1.58 \%. For comparison we
consider in the Appendix also the weak-coupling limit where the
cross section is a constant, independent of the relative momentum
of the colliding particles. In this case the leading correction is
found to increase the lower bound by 3/202 or 1.49 \% in precise
agreement with the known result for hard-sphere molecules
\cite{Chapman}.

Since medium effects are negligible for $T\gg T_F$, we conclude
that Eq.\ (\ref{etaclasslimit}) is a very accurate  expression for
$\eta$ at high temperatures in the unitarity limit. At low $T$
however, we saw by comparing (\ref{ViscFormula}) and
(\ref{etarelax}) that medium effects are important. Compared to
the Kubo result, the variational solution includes medium effects
only in the sense that the medium scattering rate
$\tau^{-1}(p)=-2{\rm Im}\Sigma(p,\xi_p)$ is used in the Boltzmann
equation implicitly assuming ${\rm Im}\Sigma\ll \epsilon_F$
whereas ${\rm Re}\Sigma$ is neglected~\cite{BruunSmith}. On the
other hand, the variational nature of the solution corresponds to
approximately summing diagrams beyond the approximation in  Fig.\
\ref{FeynFig} leading to the Kubo result (\ref{ViscFormula}).
Comparing the two approximations for $\eta$,  we see that the
variational  solution of the Boltzmann equation obtained in Ref.\
\cite{BruunSmith} agrees reasonably well with the Kubo result for
low $T$. Note that since in the unitarity limit
$T_c/\mu(T_c)\approx0.6$ is rather large, the $T^{-2}$ divergence
in $\eta$ for $T\rightarrow 0$ due to Fermi blocking is not seen
for $T\ge T_c$ in Fig.\ \ref{etaFig}.

\section{Comparison with experiment}\label{exp}

When conditions are hydrodynamic, the attenuation of a collective
mode is related to the viscosity. We now use our calculated
viscosity to interpret the measured~\cite{Kinast,Grimm}   damping
of the collective modes in an atomic gas trapped in a potential of
the form
\begin{equation}
V(x,y,z)=\frac{m}{2}(\omega_x^2x^2+\omega_y^2y^2+\omega_z^2z^2).\end{equation}
The attenuation $\Gamma$ of a collective mode is defined in terms
of the amplitude decay of the density oscillations given by one
half the rate of change of mechanical energy, i.e.
\begin{equation}
\Gamma =-\frac{\langle\dot{E}_{\rm mech}\rangle_t}{2\langle E_{\rm mech}\rangle_t},
\end{equation} where $\langle\ldots\rangle_t$
indicates the time average over a period of the
cycle~\cite{Kavoulakis}. The modes we examine are characterized by
a velocity field ${\bf v}=(ax,by,cz)$. Following
Ref.~\cite{Kavoulakis}, we obtain
\begin{equation}
\Gamma=\frac{2(a^2+b^2+c^2-ab-ac-bc)\int
d^3r\,\eta({\mathbf{r}})}{3m\int d^3r\,n({\mathbf{r}})
(a^2x^2+b^2y^2+c^2z^2)} \label{Damping}
\end{equation}
for the damping. Here $n({\mathbf{r}})$ is the  density of atoms in
the trap. In the unitarity limit, (\ref{alphadef}) gives
$\eta({\mathbf{r}})=n({\mathbf{r}})\hbar\alpha(T/T_F({\mathbf{r}}))$
with $T_F({\mathbf{r}})=[3\pi^2n({\mathbf{r}})]^{2/3}/2m$.
We parametrize  $\alpha$ for $T\ge T_c$ in the form
\begin{equation}
\alpha=\alpha_0+\alpha_{3/2}(T/T_F)^{3/2}\label{alphafit}
\end{equation}
where $\alpha_0=-0.2$ and $\alpha_{3/2}=2.77$ are numbers chosen to fit our
numerical results for $\eta$ in Sec.\ \ref{kubo}.

The $\alpha_0$-term yields a spatial integral
of $n({\mathbf{r}})$ in (\ref{Damping}) whereas the $\alpha_{3/2}$
term gives a spatial integral of a constant since the viscosity in
the classical limit is independent of density. The integration
must be cut off near the edge of the cloud where the density is
small and hence the mean free path long, resulting in the
breakdown of the validity of hydrodynamics. We adopt the procedure
described in Ref.\ \cite{Kavoulakis} that the hydrodynamic
description holds out to a distance given by the condition that an
atom incident from outside the cloud has a probability of no more
than $1/e$ of {\it not} suffering a collision. Since the density
is well approximated by its classical value near the edge we
obtain as in~\cite{Kavoulakis} a cutoff distance that depends
weakly on the cross section $\sigma$. In the unitarity limit this
cross section is $\sigma=C\lambda_T^2$, where
$\lambda_T=\sqrt{2\pi/mT}$ is the thermal de Broglie wavelength
and $C$ is a numerical constant of order unity.

In the experiments~\cite{Grimm,Kinast}, the trap is very elongated
with $\omega_z\ll \omega_x,\omega_y$. The transverse frequencies
$\omega_x$ and $\omega_y$ differ by 10-20\%, but to leading order
this ellipticity can be taken into account for the mode
frequencies by considering a cylindrically symmetric trap
\begin{equation}
V(x,y,z)=\frac{m}{2}\omega_\perp^2(x^2+y^2+\lambda^2 z^2)
\end{equation}  
with $\omega_\perp=\sqrt{\omega_x\omega_y}$. To model the experiments,
we therefore consider such a trap in the following with
$\lambda\ll 1$, which results in a separation of the transverse
and longitudinal motion. Hence, $a=b$ and $c=0$ for the transverse
mode, while $a=b=0$ and $c\neq 0$ for the axial mode.

In the classical limit with $\eta$ given by (\ref{etaclasslimit}), Eq.\
(\ref{Damping}) yields
\begin{equation}
\frac{\Gamma}{\omega_\perp}=1.08\lambda^{2/3}N^{-1/3}\left(\frac{T}{T_F}\right)^2
f(\lambda,\tau_0), \label{dampingrateclass}
\end{equation}
where $f(\lambda,\tau_0)$, given in Ref.~\cite{Kavoulakis},
 is an angular average arising from the integration over
volume,  with $\tau_0=\sigma n(0)\sqrt{T/2m}/2\lambda\omega_\perp$. The total number of
trapped atoms is $N$. For the axial mode, one obtains a similar
expression for $\Gamma/\omega_z$ with $\lambda^{2/3}$ replaced by
$\lambda^{5/3}$, while the numerical constant in front is $3.1$.

We shall now compare our calculated damping rates with experiments
on trapped $^6$Li atoms~\cite{Grimm,Kinast}. Measurements of the
damping of the axial mode~\cite{Grimm} yielded the value
$\Gamma/\omega_z\approx 0.0045$ in the unitarity limit
for a trap with
$\lambda=0.030$ and $N=4\times 10^5$. The
temperature, however, was not known, and we cannot therefore make a
direct comparison to our calculated damping. To obtain an estimate,
we use the classical result (\ref{dampingrateclass}),
which yields values in the range $0.004\le\Gamma/\omega_z\le0.007$
for $0.3\le T/T_F\le0.6$
($C=1$).

\begin{figure}
\includegraphics[width=0.8\columnwidth,
height=0.6\columnwidth,angle=0,clip=]{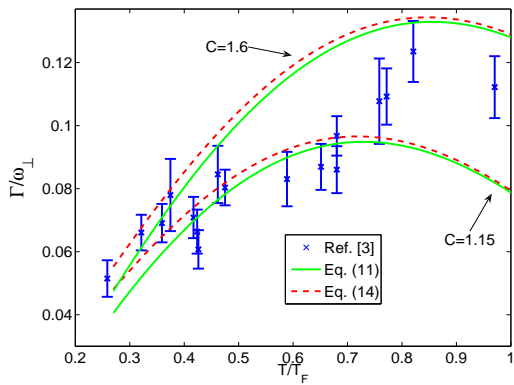} \caption{(Color
online) The damping of the transverse mode. The $\times$'s are experimental values from ~\cite{Kinast}
and lines are theory.} \label{DampFig}
\end{figure}

The Duke experiments~\cite{Kinast} allow for a more direct
comparison, since information on the temperature is available. In
Fig.\ \ref{DampFig}, we plot the observed damping rate of the lowest
transverse collective mode.
The temperature is determined by fitting the observed density
profile to an ideal gas profile with a rescaled Thomas--Fermi radius
$\tilde{R}_{TF}^2=\sqrt{1+\beta}R_{TF}^2$ where $R_{TF}^2=2(3\lambda
N)^{1/3}a_\perp^2$ and
$a_\perp^{-2}=m\omega_\perp$~\cite{Kinast}. Here, $(1+\beta)$ is
the parameter used conventionally in the unitarity limit, equal to
the ratio between the ground-state energy and that of a
non-interacting Fermi gas. In the classical regime, the fitted
temperature $\tilde{T}$ is related to the physical temperature by
$\sqrt{1+\beta}\tilde{T}=T/T_F$. We have used this relation for all
$T$ (with $\beta=-0.5$), since the
profile is approximately classical above the critical temperature~\cite{JThomasUnpub}.

We plot the calculated  damping rate from (\ref{Damping}) for two different values of $C$
of order $1$. The spatial integrals
are performed with a density profile corresponding to an ideal gas
with temperature $\tilde{T}$, Thomas--Fermi length $\tilde{R}_{TF}$,
$N=2\times10^5$,  and $\lambda=0.045$.  We also plot the classical
limit (\ref{dampingrateclass}). As expected from Fig.\
\ref{etaFig}, we see that the effects of the medium are small
except for low $T$. Note that our theoretical curves are not reliable
in the region where they predict a decreasing damping with increasing temperature,
since this reflects the breakdown of hydrodynamic behavior:
for larger $T$ the cloud becomes
less dense and the cross section $\sigma$ also decreases as $1/T$,
implying that the mean free path increases with temperature.  To
estimate the temperature above which the behavior ceases to be
hydrodynamic, we compare the mean free path $\ell(0)\sim 1/n(0)\sigma$ 
in the center to the spatial extent of the cloud in
the transverse direction. In the classical limit $\ell(0)\sim R_xR_yR_z/N\sigma$, 
where $R_i=(2\pi T/m\omega_i^2)^{1/2}$. The
condition $\ell(0)\ll R_\perp$ implies that
$T\ll\omega_\perp(\lambda N)^{1/2}$. For the experimental
parameters, this means that  the gas is hydrodynamic for
$T\ll 5\mu$K$\sim2T_F$ and there is a limited temperature
range in which we can compare the theory to the measured damping rate, since
our calculations only apply to the normal phase.
The fact that the observed mode remains hydrodynamic with a small damping for  $T\rightarrow 0$
indicates superfluidity~\cite{Kinast2}.

The results for the damping yield the correct order of magnitude
for two experiments measuring at two very different frequencies. 
This cannot be obtained simply by adjusting the parameter $C$ since the
results depend only weakly on $C$  and we have used $C\sim 1$ in
both cases.

\section{Discussion and conclusions}
The shear viscosity of a normal Fermi gas in the unitarity limit was analyzed as a function of temperature.
For high temperatures where one can perform systematic calculations, we demonstrated that a variational 
solution to the Boltzmann equation yields a value of $\eta$ which is 
accurate to better than two percent. At low temperatures, the role 
of the strong--coupling pseudogap effect was analyzed by calculating the viscosity within the Kubo formalism. We
showed that the pseudogap reduces the viscosity significantly since it suppresses the density of states 
at the Fermi level. While we stress that our
calculations of the viscosity are approximate
in nature for $T_c\le T\lesssim T_F$, it is interesting that the Kubo
approach yields values that are close to
those obtained from the Boltzmann equation (with a medium cross section) in this regime.
This suggests that the main effects of the medium can be taken into account by using a
medium cross section in a Boltzmann approach to  the transport properties of the
atomic cloud in the unitarity limit.
Based on these high and low temperature results, we constructed an interpolation formula for the 
viscosity as a function of temperature at unitarity. This formula was used to analyze two experiments 
on the decay of collective modes in terms of viscous damping. In performing this analysis, it was
crucial to introduce an explicit cut-off in the spatial integrations since hydrodynamics
necessarily fails in the outer parts of the cloud, where the density is low. 
We concluded that the hydrodynamic approach of viscous damping 
 accounts reasonably well for the experimental
observations; this holds for both  the longitudinal and the transverse modes
for which the observed damping differs by an order of magnitude. 
 It would be very
valuable to be able to compare theory and experiment  at higher
temperatures, where the behavior should be collisionless.
In this
limit  the mode frequencies should approach twice the oscillator
frequencies and the damping rate become proportional to the inverse
relaxation time rather than the viscosity.

We are grateful to J.\ E.\ Thomas for very helpful correspondence. We also
thank T.-L.\ Ho and C.\ J.\ Pethick for discussions.

\appendix
\section{The high-temperature limit}\label{hightemp}
In this appendix we derive the leading correction to the
lowest-order variational result for the shear viscosity of a
classical gas. As is well known (see e.g.\ paper I for details),
the properties of the collision integral in the Boltzmann equation
allow one to derive a lower bound on the viscosity of the form
\begin{equation}\eta\geq \frac{(U,X)^2}{(U, HU)},\label{lowerbound}\end{equation} where $(...,...)$ denotes a
suitably  defined scalar product. Here $X$ denotes the
inhomogeneous  term in the Boltzmann equation while $U$ is the
trial function and the positive, semi-definite integral operator
$H$ represents the collision term. Since we are only interested in
determining the relative correction to the high-temperature
viscosity arising from an improved trial function we use units
such that  $2m=k_BT=1$ in order to simplify the presentation. It is
convenient to consider the equivalent upper bound on the inverse
viscosity, and we shall therefore seek to minimize the functional
$F$ given by
\begin{equation}F= \frac{(U, HU)}{(U,X)^2}\label{upperbound}\end{equation}
by varying the trial function $U$. The lower bound on the
viscosity given in  (\ref{etaclasslimit}) corresponds to the
choice $U=X$.

In order to improve this bound we derive the minimum value of $F$
for a trial function $U$ given by a variable linear combination of
the functions $U_1$ and $U_2$,
\begin{equation}
U=\gamma U_1+c(1-\gamma)U_2,
\end{equation}
where $\gamma$ is a parameter to be varied, while $c$ is a
constant. We choose $c$ such that $(U,X)$ is independent of
$\gamma$,
\begin{equation}
c=\frac{(U_1,X)}{(U_2,X)},
\end{equation}
with which we  obtain $(U,X)=(U_1,X)$. The numerator in
(\ref{upperbound}) is a quadratic form in $\gamma$, which is
readily minimized resulting in the value $F_{\rm min}$. We are
interested in the relative difference between $F_{\rm min}$ and
the value $F_{\gamma=1}$ of $F$ for $\gamma=1$. Consequently we
define $\delta$ according to
\begin{equation}
\delta=1-\frac{F_{\rm min}}{F_{\gamma=1}},
\end{equation}
where $F_{\gamma=1}=(U_1,HU_1)/(U_1,X)^2$. The integral operator
$H$ is symmetric and therefore $(U_1, HU_2)=(U_2,HU_1)$. We define
the quantities $h_{12}$ and $h_{22}$ by
\begin{equation}
h_{12}=c\frac{(U_1,HU_2)}{(U_1,HU_1)}\;\;\;{\rm and} \;\;\;
h_{22}=c^2\frac{(U_2,HU_2)}{(U_1,HU_1)},
\end{equation}
in terms of which  $\delta$ assumes the form
\begin{equation}
\delta=\frac{(1-h_{12})^2}{1+h_{22}-2h_{12}}. \label{delta}
\end{equation}
Note that both $h_{12}$ and $h_{22}$ are independent of any
constant multiplying $U_1$ or $U_2$. We  take $U_1({\bf
p})=(p_z^2-p^2/3)$ (corresponding to $U_1\propto X$) and $U_2({\bf
p})=(p_z^2-p^2/3)p^2$. We also define
\begin{equation}
\Delta_i=\frac{1}{2}\left(U_i({\bf p})+U_i({\bf p}_1)-U_i({\bf
p}')-U_i({\bf p}_1')\right)
\end{equation}
for $i=1,2$. It is convenient to introduce relative and
center-of-mass momentum variables according to
\begin{equation}
{\bf p} ={\bf q}+\frac{{\bf Q}}{2},\; {\bf p}_1 =-{\bf
q}+\frac{{\bf Q}}{2};\;{\bf p}' ={\bf q}'+\frac{{\bf Q}}{2},\;{\bf
p}_1' =-{\bf q}'+\frac{{\bf Q}}{2}.
\end{equation}
Since the scattering is elastic, we have $q^2=q'^2$.  We obtain
\begin{equation}
\Delta_1=q_z^2-q_{z}'^2
\end{equation}
and
\begin{gather}
\Delta_2=(q_z^2-q_{z}'^2)\left(q^2+\frac{Q^2}{4}\right)+ {\bf
q}\cdot{\bf Q}\left(q_zQ_z-\frac{{\bf q}\cdot{\bf Q}}{3}\right)\nonumber\\
-{\bf q}'\cdot{\bf Q}\left(q_z'Q_z-\frac{{\bf q}'\cdot{\bf
Q}}{3}\right).
\end{gather}
In order to determine $h_{12}$ and $h_{22}$ we  first calculate
the angular averages $\langle\Delta_1^2\rangle$, $\langle\Delta_1\Delta_2\rangle$ and
$\langle\Delta_2^2\rangle$ by integrating over the directions of each of the vectors
$\bf q$, ${\bf q}'$ and $\bf Q$. We get
\begin{equation}
\langle\Delta_1^2\rangle=\frac{8}{45}q^4
\end{equation}
together with
\begin{equation}
\langle\Delta_1\Delta_2\rangle=\frac{8}{45}q^4\left(q^2+\frac{7}{12}Q^2\right)
\end{equation}
and
\begin{eqnarray}
\langle\Delta_2^2\rangle=\frac{8}{45}q^4\left(q^2+\frac{Q^2}{4}\right)^2 \nonumber\\
+\frac{16}{135}Q^2q^4\left(q^2+\frac{Q^2}{4}\right)
+2q^4Q^4\left(\frac{2}{75}-\frac{4}{405}\right).
\end{eqnarray}
The final integrations over the magnitude of the total and
relative momentum are given by
\begin{equation}
\int_0^{\infty}dq
q^2q\sigma(q)\int_0^{\infty}dQQ^2e^{-2q^2-Q^2/2}\langle\ldots\rangle,
\label{finint}
\end{equation}
where the exponential factors arise from the product of the
equilibrium distribution functions $f(\xi_p)f(\xi_{p_1})$ and
$\sigma(q)$ is the  $q$-dependent cross section. The additional
factor of $q$ in the integrand of (\ref{finint}) is due to the
relative velocity occurring in the collision integral. Putting
these results together and using that
$c=\Gamma(7/2)/\Gamma(9/2)=2/7$ we finally get
\begin{equation}
h_{12}=\frac{2}{7}\left(\frac{I_6}{I_4}+\frac{7}{4}\right)
\end{equation}
and
\begin{equation}
h_{22}=\frac{4}{49}\left(\frac{I_8}{I_4}+\frac{7}{2}\frac{I_6}{I_4}+\frac{301}{48}\right),
\end{equation}
where we have defined the integrals $I_n$ by
\begin{equation}
I_n=\int_0^{\infty}dq q^{n+3}\sigma(q)e^{-2q^2}.
\end{equation}
In the weak-coupling limit $\sigma$ is independent of $q$ and we
obtain $I_8/I_4=5$ and $I_6/I_4=2$, resulting in $h_{12}=15/14$
and $h_{22}=877/588$. When inserted into (\ref{delta}) these
values  yield $\delta=3/205$ in precise agreement with
\cite{Chapman}, since $(1-\delta)^{-1}= 1+3/202$. As shown in
Ref.\ \cite{Chapman} the convergence is very fast, the next-order
term yielding a further correction of only a tenth of a percent.
For resonant scattering $\sigma\propto q^{-2}$, in which case
$I_8/I_4=3$ and $I_6/I_4=3/2$, resulting in $h_{12}=13/14$ and
$h_{22}=697/588$, which yields $\delta=3/193$ or correspondingly
$(1-\delta)^{-1}=1+3/190$.

We can thus safely conclude that the expression
(\ref{etaclasslimit}) is accurate to better than two percent at
all values of $a$, including the unitarity limit
$a\rightarrow\infty$.

\end{document}